\documentclass{aa}
\usepackage{soul}
\soulregister\cite7
\soulregister\ref7
\usepackage{siunitx}
\usepackage{hyperref}
\usepackage{graphicx} 
\usepackage{makecell}
\usepackage{txfonts}
\usepackage{subcaption}         
\usepackage{lscape}             
\usepackage{placeins}           
\usepackage{booktabs}           
\usepackage{multirow}           
\usepackage{orcidlink}

\title{SMART : Spherical Mass ApeRture Toolkit}
\author{I. Bezmaternykh
        \inst{\ref{dedip}} \and
        S. Pires
        \orcidlink{0000-0002-0249-2104}
        \inst{\ref{dedip}, \ref{aim}} \and
        N. Dagoneau
        \inst{\ref{dedip}} \and
        G. W. Pratt
        \inst{\ref{aim}} \and
        L. Chappuis
        \inst{\ref{dedip}} \and
        G. Leroy
         \orcidlink{0009-0004-2523-4425}
         \inst{\ref{durham}}
        }
\date{April 2026}
   \institute{
    Université Paris-Saclay, CEA, Département d’Électronique des Détecteurs et d’Informatique pour la Physique, 91191, Gif-sur-Yvette, France
    \label{dedip}
    \and
   Université Paris-Saclay, Université Paris Cité, CEA, CNRS, AIM, 91190 Gif-sur-Yvette, France\\
   \email{sandrine.pires@cea.fr}
   \label{aim} 
   \and
   Institute for Computational Cosmology, Department of Physics, Durham University, South Road, Durham DH1 3LE, United Kingdom
    \label{durham} 
    }
\begin{document}
  \abstract
    {
 The aperture mass filter is a widely used tool in weak lensing studies. Aperture mass corresponds to the projected density field convolved by a compensated filter of a specific scale which can be measured directly from the shape of background galaxies.
While previous surveys have typically relied on the flat-sky approximation, forthcoming Stage IV surveys such as Euclid, the Vera C. Rubin Observatory or the Nancy Grace Roman Space Telescope will cover sufficiently large areas to require the construction of curved-sky aperture mass maps.
In this paper, we extend the shear-based aperture mass formalism to the sphere and introduce a software package for a fast and precise computation of spherical aperture mass maps from the shear field, defined directly on the celestial sphere. Unlike existing approaches, our method does not rely on planar projections and does not require the reconstruction of
convergence maps. The software provides four different implementations, all of which are independent of the choice of aperture mass filter function. Starting from the brute-force approach, whose computational cost renders it impractical for future surveys covering more than 10,000 deg$^2$, we investigate alternative galaxy-grouping strategies to improve computational efficiency. We demonstrate speed-up factors of up to 40 at a resolution of \texttt{NSIDE} = 4096 while achieving significantly higher precision than existing state-of-the-art methods. In particular, we show that the spherical shear-based implementations using a pixel-based approach outperforms the existing methods in terms of both precision and computational time. 
Finally, we show that, for weak-lensing cluster detection, the cluster sample produced using the spherical shear-based aperture mass method consistently outperforms those obtained using projected shear-based approaches in the purity–completeness plane, while simplifying the resulting selection function.
Our spherical mass aperture software, called SMART, is made available on \href{https://github.com/cea-lilas/SMART}{GitHub}.
}

\maketitle
\nolinenumbers
\section{Introduction}
Weak gravitational lensing, the deflection of light by matter along the line-of-sight, provides a direct probe of the matter distribution in the Universe, including dark matter, and can therefore be compared directly with theoretical models of structure formation.
Weak lensing is a statistical probe that relies on measuring the distortions of galaxy shapes (known as cosmic shear) across a large sample of galaxies. 
A widely used technique for relating the shear field to the projected mass distribution on a given angular scale is the aperture mass filter introduced in \cite{Schneider_1996}. The resulting aperture mass maps are particularly useful in cosmological analyses, especially those involving higher-order statistics \citep[e.g.][]{Ajani_2023,Vinciguerra_2026}. They are also used in galaxy cluster studies to identify clusters by detecting the lensing signal they induce on background galaxies \citep[e.g.][]{Miyazaki_2018,Hamana_2020,Oguri_2021}.

The full potential of weak gravitational lensing has been progressively unlocked thanks to wide-field optical surveys covering hundreds of square degrees, such as CFHTLenS \citep{Heymans_2012}, KiDS \citep{deJong_2013}, DES \citep{Flaugher_2005}, and HSC \citep{Aihara_2018}. Until now, the construction of aperture mass maps from shear measurements in observational data has typically relied on the flat-sky approximation. 
In practice, the improvement gained by moving from a flat-sky to a curved-sky treatment is marginal when the survey area is of order $\sim 100 \deg^2$ \citep[e.g.][]{Wallis_2017,Pires_2020}. 
However, upcoming wide weak lensing surveys will probe thousands of square degrees, including Euclid \citep{Laureijs_2011}, the Vera C. Rubin Observatory \citep{Ivezic_2019}, and the Nancy Grace Roman Space Telescope \citep{Spergel_2015}. With the increasing sky coverage of current and forthcoming surveys, a rigorous treatment of weak lensing data directly on the celestial sphere becomes necessary.

The formalism for reconstructing curved-sky convergence maps from shear maps (known as mass inversion) has already been introduced \citep[e.g.][]{Wallis_2017,Chang_2018} and aperture mass filters can also be applied to these reconstructed convergence maps to produce aperture mass maps on the sphere.
However, this approach is less precise than the shear-based approach because the shear signal is averaged over pixels before performing the mass inversion, meaning the accuracy of the aperture mass maps depends on the pixel size. 
Additionally, the mass inversion itself introduces biases, notably due to masking. These effects can be partially mitigated by applying corrections for mass mapping systematic effects \cite[e.g.][]{Pires_2020} but this significantly increases the computational time of the mass inversion.
An even more critical limitation is that applying the aperture mass filter to convergence maps does not provide a straightforward way to estimate the associated noise. In this case, the noise in the aperture mass maps can only be assessed by randomly rotating the orientations of source galaxies, then performing the mass inversion and applying the aperture mass filter to the resulting convergence maps.
This procedure must be repeated N times (with N>100), making it computationally costly.
This becomes particularly prohibitive when using mass inversion methods that include corrections for systematic effects.

For this reason, the shear-based mass aperture approach is largely preferred, but for wide-field surveys, standard approaches rely on planar projections : either by projecting the entire fields \citep[e.g.][]{Oguri_2021} or on the subdivision of a large-scale survey into small patches \citep[e.g.][]{Martinet_2018,Harnois_Deraps_2024}, both of which are not fully satisfactory. A fully consistent formulation of the shear-based mass aperture approach on the sphere is, however, still lacking in the literature. 

The aim of this article is to introduce the spherical formalism for the shear-based mass aperture, defined directly on the celestial sphere without relying on planar projections or requiring the reconstruction of convergence maps. We present several numerical implementations of the estimator and discuss their respective computational properties. The precision and accuracy of the codes, together with their computational efficiency, are tested using a mock galaxy catalogue specially built for this purpose. 
As an illustration, we apply this framework to the problem of galaxy cluster detection 
to highlight the relevance of the spherical approach for wide-field surveys. 
The associated software implementation
are made freely available on \href{https://github.com/cea-lilas/SMART}{GitHub}\footnote{https://github.com/cea-lilas/SMART}.

In Section~\ref{sec:formalism}, we introduce the standard Cartesian formulation of the aperture mass. We then discuss existing approaches for handling large-field surveys, highlighting their limitations. Finally, we derive the corresponding spherical formulation, providing a consistent extension of the Cartesian formalism to the sphere.
In Section~\ref{sec:implementation}, we present four different implementations of the spherical shear-based aperture mass estimator for the construction of full-sky maps using the HEALPix pixelisation scheme.
Section~\ref{sec:data} describes the simulated galaxy catalogue used throughout this work. 
In Section~\ref{sec:evaluation}, we assess the precision and computational performance of the different implementations of the spherical shear-based mass aperture with respect to current approaches. 
In Section~\ref{sec:application}, we apply the methods to the problem of weak-lensing galaxy cluster detection and compare their performances in terms of purity and completeness. 
Finally, our conclusions are presented in Section~\ref{sec:conclusion}.

\section{Formalism}
\label{sec:formalism}

To study the matter distribution, dominated by dark matter, it is necessary to account for the noise caused by the random distribution of intrinsic galaxy shapes. For this purpose, the aperture mass filter has been introduced by \cite{Schneider_1996} and is commonly used to reconstruct maps of the projected matter distribution at a given scale, referred to as aperture mass maps. The aperture mass can be computed from the convergence field $\kappa$ by integrating the convergence within an aperture filter $U$ centred at position $\theta_0$ such that :
\begin{equation}
    M_{\rm ap}(\boldsymbol{\theta_{\rm o}}) = \int_{\mathbb{R}^2} \kappa(\boldsymbol{\theta})\, U(\lvert\, \boldsymbol{\theta} \, -\, \boldsymbol{\theta_{\rm o}} \,\rvert)\, d^2 \boldsymbol{\theta}, 
    \label{Eqn:Map}
\end{equation}
where $U$ must be compensated; that is :
\begin{equation}
    \int_{0}^{\rm \theta_{\rm ap}}\rm \theta \, U(\rm \theta)\,d\theta = 0,
\end{equation}
must be fulfilled in the aperture of radius $\theta_{\rm ap}$.

The aperture mass maps can also be computed from the tangential and cross shear $\gamma_{+}$ and $\gamma_{\times}$ by applying the aperture filter $Q$, as follows:
\begin{equation}
    \begin{aligned}
        M^E_{\rm ap}(\boldsymbol{\theta_o})= \int_{\mathbb{R}^2} \gamma_{+} (\boldsymbol{\rm \theta})\, Q(\lvert\, \boldsymbol{\rm \theta} \, -\, \boldsymbol{\rm \theta_{\rm o}} \,\rvert)\, d^2 \boldsymbol{\rm \theta}, \\
        M^B_{\rm ap}(\boldsymbol{\theta_o})= \int_{\mathbb{R}^2} \gamma_{\times} (\boldsymbol{\rm \theta})\, Q(\lvert\, \boldsymbol{\rm \theta} \, -\, \boldsymbol{\rm \theta_{\rm o}} \,\rvert)\, d^2 \boldsymbol{\rm \theta}, 
    \end{aligned}
    \label{map_gammat}
\end{equation}
with : 
\begin{equation}
    \begin{aligned}
        \gamma_{+} &=& \gamma_1 \cos (2 \psi) + \gamma_2 \sin (2 \psi), \\
        \gamma_{\times} &=& -\gamma_1 \sin (2 \psi) + \gamma_2 \cos (2 \psi),
    \end{aligned}
\label{gammat}
\end{equation}
where $\gamma_1$ and $\gamma_2$ are the two components of the shear and $\psi$ is the polar angle $\psi(\boldsymbol{\theta}, \boldsymbol{\theta_0)}$ relative to the centre of the aperture $\boldsymbol{\theta_0}$\\

\noindent and:
 
\begin{equation}
Q(\theta) = \frac{2}{\theta^2}\int_0^{\theta}{\theta'U(\theta')d\theta'-U(\theta)}.
\end{equation}


\begin{figure*}[t]
    \centering
    \hspace{0.4cm}
    \begin{subfigure}{0.38\textwidth}
        \centering
          \includegraphics[width=\linewidth]{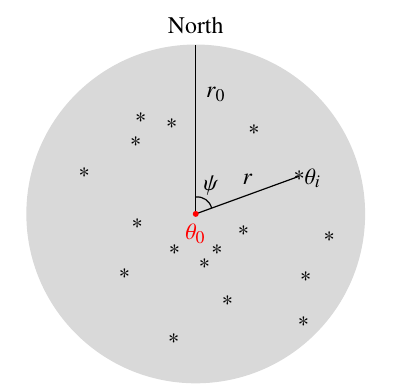}
        \caption{\footnotesize Cartesian approach}
        \label{fig:cart_angles}
    \end{subfigure}
\hspace{1.7cm}
    \begin{subfigure}{0.35\textwidth}
        \centering
       \includegraphics[width=\linewidth]{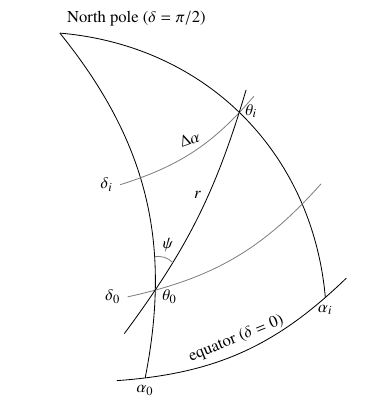}
        \caption{\footnotesize Spherical approach}
    \end{subfigure}
    \caption{\footnotesize \footnotesize Application of a mass aperture filter centred at position $\theta_0$ to a galaxy at position $\theta_i$, given the angular distance $r$ and the polar angle $\psi$. \textit{(Left)} Cartesian projection. The filter aperture or radius $r_0$ is coloured in gray. Asterisks ($\ast$) represent galaxies. \textit{(Right)} Spherical representation. Great circles are drawn in solid black.
    }
        \label{fig:sph_angles}
\end{figure*}

\subsection{Spherical convergence-based approach}
\label{sec:conv}
In earlier weak lensing studies \citep[e.g.][]{Marian_2012, Pires_2012}, the convergence was frequently used to reconstruct aperture mass maps. 
However, recent studies tend to work directly with the shear field \citep[e.g.][]{Martinet_2015, Hamana_2020, Harnois_Deraps_2024}. This shift is motivated by the fact that the shear (or more exactly the reduced shear) is a direct observable and is therefore usually preferred for simplicity reasons. In particular, the shear-based approach avoids the mass inversion step required to reconstruct convergence maps, thereby reducing the systematic effects introduced by this procedure. Moreover, it provides a more accurate estimation, as the filter weight $Q$ can be evaluated directly at the actual positions of the galaxies rather than at pixel centres as is done on convergence-based approaches. Furthermore, the aperture mass filter applied to the convergence maps does not allow for a direct estimation of the associated noise maps.

Although not optimal, the formalism for reconstructing curved-sky convergence maps from shear maps has already been established \citep[e.g.][]{Wallis_2017,Chang_2018}. Consequently, spherical aperture mass maps can be readily obtained by applying the aperture filter $U$ to the spherical convergence maps $\kappa_E$ and $\kappa_B$, as follows: 
\begin{equation}
    \begin{aligned}
        M^E_{\rm ap}(\boldsymbol{\theta_{\rm o}}) = A \sum_i \kappa_E(\boldsymbol{\theta_i})\, U(\lvert\, \boldsymbol{\theta_i} \, -\, \boldsymbol{\theta_{\rm o}} \,\rvert)\ \\
        M^B_{\rm ap}(\boldsymbol{\theta_{\rm o}}) = A \sum_i \kappa_B(\boldsymbol{\theta_i})\, U(\lvert\, \boldsymbol{\theta_i} \, -\, \boldsymbol{\theta_{\rm o}} \,\rvert)\ 
        \label{Eqn:Map}
    \end{aligned}
\end{equation}
where the sum $i$ runs over all the pixels with centres located inside the aperture, and $A$ denotes the surface area of a pixel in arcmin$^2$.



\subsection{Cartesian (projected) shear-based approach}
\label{sec:proj}
To date, owing to the relatively small sky coverage involved, the construction of shear-based aperture mass maps has been undertaken assuming the flat-sky approximation using a Cartesian approach \citep{Martinet_2018,Oguri_2021,Harnois_Deraps_2024}.
In these studies, the mass aperture is computed by integrating the shear within the aperture centred at the pixel position $\theta_0$, with each galaxy contribution weighted by the value of the filter function $Q$ evaluated at the galaxy position relative to the aperture centre.
Since the shear is only measured at the locations of source galaxies, the continuous integral over the aperture is replaced by a discrete sum over galaxies. 
This sum is weighted by the galaxy density per arcmin$^2$ ($n_{\rm gal}$): 
\begin{equation}
    \begin{aligned}
        M^E_{\rm ap}(\boldsymbol{\theta_{\rm o}}) = \frac{1}{n_{\rm{gal}}} \sum_i \gamma_+(\boldsymbol{\theta_i})\, Q(\lvert\, \boldsymbol{\theta_i} \, -\, \boldsymbol{\theta_{\rm o}} \,\rvert)\ \\
        M^B_{\rm ap}(\boldsymbol{\theta_{\rm o}}) = \frac{1}{n_{\rm{gal}}} \sum_i \gamma_\times(\boldsymbol{\theta_i})\, Q(\lvert\, \boldsymbol{\theta_i} \, -\, \boldsymbol{\theta_{\rm o}} \,\rvert)\ 
        \label{Eqn:map}
    \end{aligned}
\end{equation}
where the sum $i$ runs over all the pixels in the aperture.
In Sect.~\ref{sec:SUMvsAVG}, we explain how the galaxy density ($n_{\rm gal}$) is computed.
Shear-based calculations are particularly convenient for applications that require computing of the variance of aperture mass maps arising from the intrinsic ellipticity of the source galaxies $\sigma_{\epsilon}$, since the variance can be evaluated directly as follows:

\begin{equation}
    \sigma^2_{\rm ap}(\boldsymbol{\theta_o})=\frac{1}{2 n_{\rm{gal}}^2} \sum_i
    \gamma^2 (\boldsymbol{\rm \theta_i})\, Q^2(\lvert\, \boldsymbol{\rm \theta_i} \, -\, \boldsymbol{\rm \theta_{\rm o}} \,\rvert).
    \label{Eqn:map_noise}
\end{equation}
In this framework, the distances and angles appearing in Eq.~\ref{gammat}, Eq.~\ref{Eqn:map} and Eq.~\ref{Eqn:map_noise} are defined in the Cartesian plane as illustrated in the left panel of Fig.~\ref{fig:sph_angles}.\\

\subsubsection{Projection of the position of the galaxies}
In the literature, it is common to assume the flat-sky approximation when constructing aperture mass maps \cite[e.g.][]{Oguri_2021,Harnois_Deraps_2024}, as this simplifies the computation of the angular separation $\psi$ and the distance $|\theta_i - \theta_0|$. In these studies, Cartesian formulations of the shear-based mass aperture filter are used, with distances and angles defined in the projected plane. So, prior to applying the filter, 
the shear field is projected by mapping the galaxy positions from the celestial sphere, $(\alpha_i, \delta_i)$, using a tangent plane projection (also known as gnomonic projection) centred on a reference point $\theta_c(\alpha_c, \delta_c)$, where $x_i$ and $y_i$ denote the projected Cartesian coordinates corresponding to the angular position $\theta_i$:
\begin{equation}
    \begin{aligned}
    x_i = \frac{\cos(\delta_i)\,\sin(\alpha_i - \alpha_c)}{\sin(\delta_i) \sin(\delta_c) + \cos(\delta_i) \cos(\delta_c) \cos(\alpha_i - \alpha_c)}\\
    y_i = \frac{\cos(\delta_c)\,\sin(\delta_i) - \sin(\delta_c)\,\cos(\delta_i)\,\cos(\alpha_i - \alpha_c)}{\sin(\delta_i) \sin(\delta_c) + \cos(\delta_i) \cos(\delta_c) \cos(\alpha_i - \alpha_c)}
     \end{aligned}
\end{equation}

\subsubsection{Projection of the shear}
An aspect that is not always taken into account in these studies is that projecting a spin-2 field, such as the shear field, requires projection not only of the galaxy positions but also of the shear itself. In practice, this involves applying a rotation to the shear to account for the change in the local coordinate basis induced by the projection. The dominant effect arises from the fact that the projected meridians are not parallel, so that the North direction varies across a single projected field of view. Consequently, if this rotation is not applied, the projection introduces artificial B-modes into the resulting shear field, with an amplitude that increases with latitude. This effect can be easily mitigated by defining the shear orientation with respect to the local North direction. 
In \cite{Pires_2020}, it was shown that applying this correction is sufficient to make the residual projection errors negligible compared to other sources of systematic uncertainties.

\subsection{Spherical shear-based approach}

Recently, the increasing size of surveys dedicated to weak lensing analysis, now covering hundreds of square degrees of sky, has made the errors induced by the flat-sky approximation significant and they are now no longer negligible \cite[e.g.][]{Wallis_2017, Pires_2020}. 
The mass aperture filtering can still be performed in the plane \cite[e.g.][]{Oguri_2021} by neglecting the errors introduced at the border of the field or dividing the survey into small fields \cite[e.g.][]{Harnois_Deraps_2024}. Although the flat-sky approximation remains valid in the latter approach, it is not particularly suited for large-scale analysis, and even for small-scale studies this treatment requires the recombination of the different fields and accounting for the associated border effects.
The increasing size of the survey therefore motivates the development of curved-sky treatments.

In this paper, we extend the shear-based mass aperture formalism to the sphere.
To this end, we define the distances and angles on the celestial sphere to compute the aperture filter $Q$ and the tangential and cross shear $\gamma_+$ and $\gamma_{\times}$ as illustrated in the right panel of Fig.~\ref{fig:sph_angles}.
For a galaxy located at angular position $\theta_i$, defined by its right ascension $\alpha_i$ and declination $\delta_i$, we can compute the great circle separation angle $|\theta_i - \theta_0|$ as well as the polar angle $\psi$ relative to the centre of the filter $\theta_0(\alpha_0, \delta_0)$ as follows : 
\begin{equation}
    \cos(\theta_i - \theta_0) = \sin(\delta_i) \sin(\delta_0) + \cos(\delta_i) \cos(\delta_0) \cos(\alpha_i - \alpha_0)
\end{equation}

\begin{equation}
    \tan (\psi) = \frac{\sin(\alpha_i - \alpha_0)}{\cos(\delta_i) \tan(\delta_0) -  \sin(\delta_i) \cos(\alpha_i - \alpha_0)}
\end{equation}
Then, the mass aperture and its associated variance per pixel can be computed on the curved-sky using Eqs.~\ref{Eqn:map} and \ref{Eqn:map_noise}, provided that the distances and angles are defined consistently on the celestial sphere.

\subsection{Summed vs averaged approaches}
\label{sec:SUMvsAVG}
In the literature, we find two different implementations of the aperture mass maps, depending on the way the galaxy density per arcmin$^2$ ($n_{\rm gal}$) is computed. The galaxy density per arcmin$^2$ can be measured within the aperture or within the pixel. We will refer to these approaches as the "summed" and "averaged" approaches as defined in \cite{Leroy_2023}: 
\begin{itemize}
\item The {\bf summed approach} consists of summing the values of the mass aperture filter applied to the shear for each galaxy falling in the aperture,  and then normalising by the mean galaxy density per arcmin$^2$, within the aperture. 
This approach is widely used in cosmological analyses involving peak statistics \citep[e.g.][]{Martinet_2018,Harnois_Deraps_2024} as well as for weak lensing cluster detection \cite{Leroy_2023}.
\item The {\bf averaged approach} consists of summing for each galaxy falling in the pixel, the values of the mass aperture filter applied to the shear and then normalising by the effective galaxy density per arcmin$^2$ within the pixel of the final map instead of the average galaxy density within the aperture. The averaged formulation follows more directly from the mathematical definition and is the direct analogue of the aperture mass computed from convergence maps \citep[e.g.][]{Marian_2012, Leonard_2012, Pires_2012}. This is because the convergence is constructed from the shear map, which is obtained by summing the shear contributions of galaxies within each pixel and normalising by the effective galaxy density per pixel.
\end{itemize}
Neither approach is inherently preferred, as the choice depends on the application.
The summed approach introduces correlation between the signal and the galaxy distribution, whereas the averaged approach more closely traces the projected matter distribution but in this case, it is the noise that is correlated with the galaxy distribution. 

Both implementations are available in the SMART software. The execution time and precision presented in Sect.~\ref{sec:evaluation} are only marginally affected by the choice between them.
In Sect.~\ref{sec:comp_conv}, the comparison between the spherical shear-based aperture mass and the spherical convergence-based aperture mass map is performed using an averaged approach, since the summed approach cannot be readily extended to the convergence-based method. 
Except for this comparison, the analyses presented in this paper consistently adopt the summed approach, in line with the results of \cite{Leroy_2023}, since the focus of the paper is weak lensing cluster detection.

\section{Spherical shear-based mass aperture implementation}
\label{sec:implementation}

For data defined on the celestial sphere, HEALPix \citep{Gorski_2005} has emerged as the preferred pixelization scheme for many astrophysical studies.
HEALPix partitions the sphere into a set of diamond-shaped pixels arranged to cover the entire sky with equal areas. 
This equal-area property makes the scheme particularly well suited for statistical analyses, as each pixel samples the same solid angle on the sky.
Another advantage of HEALPix is the availability of well-maintained libraries in common programming languages (i.e. Python, C++...).
These libraries provide efficient implementations of the standard operations on a HEALPix pixel map (i.e. querying pixels in a disk around a point, converting between pixel indices and spherical coordinates and vice-versa). This makes HEALPix both easy to use and computationally efficient for large-scale astrophysical analyses.

The spherical shear-based aperture mass map is thus constructed as a HEALPix map where the value assigned to each pixel is obtained by placing the filter $Q$ at the centre of that pixel.
Subsections~\ref{sec:full-cat} to \ref{sec:binned-cat}, describe methods for calculating spherical shear-based aperture mass HEALPix maps, ranging in computational complexity and accuracy. For these different methods, the associated noise can be directly computed using the same procedure based on Eq.~\ref{Eqn:map_noise}.
Note that the implementations presented in this section are following the summed approach.
\subsection{Galaxy position-based approach: brute force}
\label{sec:full-cat}
The most basic approach to the creation of a spherical shear-based aperture mass map consists of applying the aperture mass filter directly at the position of the galaxies in the catalogue.
To compute each pixel of the output map, the following steps are applied:

\begin{itemize}
    \item[1.] Galaxies located within the aperture radius from the pixel centre are selected.
    \item[2.]  For each of these galaxies, the two shear components $\gamma_1$ and $\gamma_2$ are converted into tangential and cross shear, $\gamma_+$ and $\gamma_\times$.
    \item[3.]  The value of the filter at the position of the galaxies is calculated from their distances to the pixel centre.
    \item[4.]  The value of the filter is applied to the tangential and cross shear, and the resulting values are summed over the aperture area.
    \item[5.]  The final value assigned to each pixel is then obtained by normalising this sum by the mean galaxy density per arcmin$^2$, measured within the aperture. In the absence of missing data, this is equivalent to dividing by the number of galaxies in the aperture and multiplying by the aperture area in arcmin$^2$.
\end{itemize}
This approach has two main bottlenecks in terms of computational time. The first bottleneck resides in the fact that step 1 theoretically requires to compare the pixel's centre position against all galaxy positions in the dataset. 
This can be avoided by building a KDtree from the positions of galaxies. A KDtree is a data structure that leverages the binary tree structure to allow the retrieval of a point's neighbors in logarithmic time. This is achieved by recursively splitting the space along hyperplanes that split the points at each node into two halves of equal population until all points are in separate leaves of the binary tree.
The second bottleneck is inherent to the fact that this approach treats each galaxy individually, resulting in a computational complexity that is at least linear with respect to the total number of galaxies in the dataset.

\subsection{Pixel-based approach}
\label{sec:pix-map}
In order to address the second bottleneck, a common approach, 
is to bin the galaxy catalogue into a HEALPix map. The shear measurements of galaxies falling within the same pixel are then summed to construct the observed shear maps.
The algorithm is implemented as follows:
\begin{itemize}
    \item[1.]  Create a regular grid over the full sphere. We use HEALPix pixels, which provides an equal-area partition of the spherical surface, as described above.
    \item[2.]  For each grid pixel, the ellipticity components per pixel $\gamma_1$ and $\gamma_2$ are computed as the sums of $\gamma_1$ and $\gamma_2$ for all galaxies contained within the pixel. These summed ellipticities are then normalised by the mean galaxy density per arcmin$^2$ within the aperture. The mean galaxy density per arcmin$^2$ is calculated as the mean galaxy density per pixel multiplied by the pixel area in arcmin$^2$.
    \item[3.]  The estimated ellipticity $\gamma_1$ and $\gamma_2$ per pixel is then transformed into tangential and cross shear components $\gamma_+$ and $\gamma_\times$.
    \item[4.]  The value of the filter at the position of the pixels within the aperture is calculated from its distance to the pixel centre.
    \item[5.]  The final value assigned to each pixel is obtained by applying the aperture mass filter to the pixels within the aperture, rather than to individual galaxies. Specifically, the filter is evaluated at the centre of each pixel within the aperture and multiplied by the corresponding $\gamma_+$ and $\gamma_\times$ values. These weighted contributions are then summed over all pixels within the aperture.
\end{itemize}
This approach greatly reduces the number of points onto which the aperture mass filter has to be applied, as the pixelated grid is typically set to have around 20 to 100 galaxies per pixel, addressing the second bottleneck of the brute force approach.
Additionally, a regular grid allows for much faster neighbour retrieval, given that the HEALPix "Ring" ordering scheme can leverage the fact that pixels are arranged into ordered iso-latitude rings. Thanks to this property, retrieving galaxies in a disk comes down to a 1D interval problem, the computational time of which increases linearly in the number of pixels within the aperture. So, at a given HEALPix resolution, the computational time only depends on the filter radius. This allows to fully address the first bottleneck of the brute force approach, making this method particularly fast.
However, binning pixel values evidently comes with precision drawbacks. Firstly, averaging values smooths the resulting signal, ignoring information provided by the variation of ellipticities within each pixel. Secondly, this binning takes the centre of the pixel as the position of the ellipticities, which is arbitrary and ignores the real positions of underlying galaxies and their distribution within the pixel's area.

\subsection{Pixel-based approach using galaxy position barycentre}
\label{sec:acc-pix}
In order to address this last point of pixel positions, a simple approach consists in calculating each pixel's barycentre. This barycentre can then be used as the position of the pixel when applying the aperture mass filter.
This approach allows to actually take into account galaxy positions and better reflects the underlying distribution of ellipticities.
Additionally, it does not require changing the algorithm structure, just adding an extra barycentre calculation step, which is of a negligible complexity next to the filter application loop, as it is done with a simple loop that averages RA and Dec over pixels.
Overall, this approach allows for the best precision while maintaining low time complexity thanks to the binning. However, there is necessarily a loss in precision due to the averaging of the shear signal within each pixel.

\subsection{Galaxy-position based approach: HEALPix-Based catalogue Partitioning}

\label{sec:binned-cat}
As discussed in Method~\ref{sec:pix-map}, the HEALPix-based binning addresses the two bottlenecks encountered when applying the aperture mass filter using the brute force approach. However, by binning galaxies into pixels in order to avoid iterating filter application over all galaxies individually, some precision must be lost.
So an alternative solution to this problem is to leverage the properties of HEALPix pixels for neighbor retrieval, while keeping all galaxy ellipticities.
This means dividing the sphere into a HEALPix grid and creating, for each pixel, a sub-catalogue containing all galaxies that fall within that pixel, which can then be easily accessed using the pixel number.
In this case, the ellipticity and position of each galaxy are not averaged, but stored in the sub-catalogues, and the grouping allows for faster neighbor access.
This catalogue binning can be achieved in several ways, ranging from an array of galaxy dictionaries to a 3 dimensional array, with one dimension for pixels, one for individual galaxies within each pixel and the third one for catalogue columns.
These are all however computationally expensive to produce and require both memory reallocation and additional memory space.
A more elegant approach involves ordering the galaxy catalogue so that galaxies falling into the same pixel are on continuous ranges of catalogue rows. The method is implemented as follows:
\begin{itemize}
    \item[1.]  Associate with each galaxy position the index of the HEALPix pixel in which it falls.
    \item[2.]  Sort the catalogue by pixel index so that galaxies falling into the same pixel follow each other in continuous ranges.
    \item[3.]  Create an index table that associates each pixel index with the catalogue position of the first galaxy that falls within it.
\end{itemize}
This means that the galaxies falling within the pixel $n$ can be retrieved by selecting all galaxies in the catalogue whose indices lie between $(n)$ and $(n+1)$ in the offset table.
Since these index ranges are continuous the retrieval time does not increase with the number of galaxies within each pixel.

\begin{table*}[htbp]
\centering
\renewcommand{\arraystretch}{1.2}
\begin{tabular}{llll}
\toprule
\textbf{Mass aperture method} &
\textbf{Implementation} &
\textbf{Label} &
\textbf{Description}\\
\midrule

Cartesian (projected) shear-based
& Galaxy position-based (brute force)
& \texttt{projS\_pos} & \texttt{Sect.~\ref{sec:proj2}}\\
\midrule

Spherical convergence-based
& Pixel-based (HEALPix centres)
& \texttt{sphC\_pix}  & \texttt{Sect.~\ref{sec:conv2}}\\
\midrule

\multirow{4}{*}{Spherical shear-based}
& Galaxy position-based (brute force)
& \texttt{sphS\_pos}  & \texttt{Sect.~\ref{sec:full-cat}}\\

& Pixel-based (HEALPix centres)
& \texttt{sphS\_pix}  & \texttt{Sect.~\ref{sec:pix-map}}\\

& Pixel-based (galaxy position barycentres)
& \texttt{sphS\_pixB}  & \texttt{Sect.~\ref{sec:acc-pix}}\\

& Galaxy position-based (HEALPix catalogue partitioning)
& \texttt{sphS\_posB}  & \texttt{Sect.~\ref{sec:binned-cat}} \\

\bottomrule
\end{tabular}
\caption{\footnotesize Summary of the implemented mass aperture methods.}
\label{tab:methods}
\end{table*}

\section{Data}
\label{sec:data}
To evaluate the performance of the different methods, it is necessary to have a realistic mock full-sky galaxy catalogue that include galaxy positions, shear measurements, and redshifts, as well as the corresponding halo catalogues.
In this section, we describe the construction of this simulated galaxy catalogue obtained using the \citet{Takahashi_2017} simulation and mimicking the data characteristics of the Hyper Suprime-Cam Subaru Strategic Program (HSC-SSP) \citep[][]{Aihara_2018}.

\subsection{Hyper Suprime-Cam data}
\label{sec:HSC}

We use the HSC-SSP shape catalogue to assign a spatial galaxy distribution to the galaxy sources in the simulated full-sky galaxy catalogue developed for this study and described in the following section.
The HSC-SSP is a deep, wide-field imaging survey conducted with the Hyper Suprime-Cam on the Subaru Telescope, designed to map a large fraction of the extragalactic sky with high image quality and depth. Specifically, we use the HSC Year 3 (HSC-Y3) shape catalogue based on the internal S19A data release, which contains observations collected between March 2014 and April 2019 as part of the HSC-SSP survey \citep{Miyatake_2018, Li_2022}. 


\subsection{N-body simulation}
We produce a simulated full-sky galaxy catalogue using the full-sky gravitational lensing simulation generated using multiple-lens plane ray-tracing through high-resolution cosmological N-body simulations published by \citet{Takahashi_2017}.
The datasets include full-sky maps for the shear, density, and convergence from redshifts z = 0.05 to 5.3, but they do not include the galaxy catalogue.
The pixelization of these full-sky maps follows the HEALPix pixelisation and they are given for different resolutions : \texttt{NSIDE}~=~4096, 8192 and 16384 which correspond to angular pixel sizes of 0.86, 0.43 and 0.21 arcmin. 

\subsection{Mock galaxy catalogue}


The \citet{Takahashi_2017} dataset consists of 38 files. Each file includes shear, galaxy density, and convergence maps at redshifts spanning the range z = 0.05 to 5.3. To construct a simulated full-sky galaxy catalogue, we use the maps at the highest available resolution ($\texttt{NSIDE}$ = 16384) for which each file has the size of approximately 50 GB. Due to computational and memory constraints, our analysis has been restricted to a subset of these files. For weak lensing cluster detection, a realistic source redshift distribution is not essential. As shown in \cite{Chappuis_2026}, incorporating individual source galaxy redshift information does not significantly improve detection performance. Instead, highest performance is achieved by combining all source galaxies with redshift larger than 0.4 into a single redshift bin. We therefore select four of these files with redshifts between z = 0.9 and 1.1, corresponding to a median redshift of z = 0.9479, to be representative of an Euclid-like wide field survey.
Each pixel of these shear maps is then treated as a galaxy. However, only a subset of pixels was retained to construct the galaxy catalogue.

The full-sky galaxy catalogue is constructed by selecting the pixels (treated as individual galaxies) such that the resulting galaxy density is approximately 30 gal/arcmin$^2$ as expected for Euclid and Vera C. Rubin wide survey, while reproducing a realistic spatial galaxy distribution inferred from the HSC data.
In practice, we estimate this spatial galaxy distribution by computing the fraction of galaxies expected to fall in each bin of the convergence probability distribution derived from the HSC data, described in Sect.~\ref{sec:HSC}. This procedure produces a source galaxy distribution that is correlated with the underlying mass distribution and results in an effective galaxy density of approximately 27 gal/arcmin$^2$.




\subsection{Mock dark matter halo catalogue}
\label{mock}
A dark matter halo catalogue is also provided as part of the datasets published by \citet{Takahashi_2017}. The halos were identified in the N-body simulation using the publicly available ROCKSTAR code based on the friends-of-friends algorithm \citep{Behroozi_2013}. Halo masses were computed using the spherical overdensity method, centred on each identified structure. The resulting halo population was validated by comparing its halo mass function with the analytical model of \cite{Tinker_2008}, showing good agreement. As proposed in \cite{Leroy_2023}, we only consider the halos for which the S/N is expected to be greater than two in Sect.~\ref{sec:application}.

\section{Evaluation of algorithm performance}
\label{sec:evaluation}
In this Section, we evaluate the performance of the different spherical shear-based mass aperture implementations presented in this work, and compare them with existing approaches: the Cartesian shear-based mass aperture method after tangent projection and the spherical convergence-based mass aperture method.
In Table~\ref{tab:methods} are presented the different methods and implementations that are compared in this section, together with the labels used to refer to them and the section in which each implementation is described.
The performance tests are carried out using approximately one fifth of the full-sky galaxy catalogue introduced in Sect.~\ref{mock} with a non-uniform galaxy density of 27 gal/arcmin$^2$. To simplify the evaluation of algorithm performance in terms of radial error, we consider a circular field with a radius of 50$^\circ$, covering an area of 7372.47 deg$^2$. Although the performance is not expected to depend on the choice of filter function, a specific filter must be selected for comparison purpose. We therefore adopt the filter proposed in \cite{Jarvis_2004}, which targets objects at a scale of 4$\arcmin$.

\subsection{Comparison of the 4 different spherical shear-based mass aperture implementations}
\label{sec:comp_methods}

We first compare the performance of the different spherical shear-based mass aperture implementations presented in Table~\ref{tab:methods} and described in Sect.~\ref{sec:implementation} in terms of computing time. 


\vspace{0.4cm}
\begin{table}[t!]
\centering
\renewcommand{\arraystretch}{1.2}
\begin{tabular}{lcc}
\toprule
\textbf{Method} & \textbf{$\texttt{NSIDE}$ = 2048} & \textbf{$\texttt{NSIDE}$ = 4096} \\
\midrule
\texttt{sphS\_pos}  & 18 h 06 min 42 s & 76 h 43 min 29 s \\
\texttt{sphS\_pix}  & 11 min 13 s      & 1 h 49 min 25 s \\
\texttt{sphS\_pixB} & 11 min 27 s      & 1 h 52 min 12 s \\
\texttt{sphS\_posB} & 8 h 15 min 20 s  & 39 h 28 min 10 s \\
\bottomrule
\end{tabular}
\caption{\footnotesize Execution time of the four spherical shear-based implementations described in Sect.~\ref{sec:implementation} for two \texttt{NSIDE} resolutions, which correspond to angular pixel sizes of 1.72$\arcmin$ and 0.86$\arcmin$ respectively. The reported execution times include the computation of both the aperture mass maps and the associated noise maps.}
\label{tab:comp-time-full-cat}
\end{table}

In Table~\ref{tab:comp-time-full-cat}, we report the computational time required to generate the four different spherical shear-based aperture mass maps and the associated noise maps at a resolution of \texttt{NSIDE}~=~2048 and 4096 
using the different implementations described in Sect.~\ref{sec:implementation}. 
The computing time has been estimated in a 2 × 4.42 GHz AMD EPYC 9335 32-Core processors, using 20 CPU cores and 200 GB of allocated RAM.
For a galaxy catalogue covering one fifth of the full sky, with a galaxy density of 27 gal/arcmin$^2$, the computing time of the galaxy position based approach using brute force is about 18h at a resolution of \texttt{NSIDE}~=~2048. Increasing the resolution from \texttt{NSIDE}~=~2048 to 4096 introduces a factor of 4 increase in the computational time due to the corresponding increase in the number of pixels. Additionally, the computational time of the brute force implementation also increases in $\theta(n\log(n))$ with the number $n$ of galaxies in the catalogue, making this approach prohibitive for full-sky galaxy catalogues.
For galaxy position based approaches, we find that partitioning the catalogue provides a speed-up by a factor of 2 for both \texttt{NSIDE} resolutions. This speed-up is expected to increase for larger catalogues, as the benefits of partitioning become more significant.
Moving to a pixel-based approaches reduce the computational time by factors of 100 and 40 for \texttt{NSIDE}~=~2048 and 4096, respectively, compared to the brute force implementation. The computation of the barycentre has a negligible impact on the overall runtime. 
For pixel-based approaches, we find that increasing the resolution from \texttt{NSIDE}~=~2048 to 4096 introduces a factor of approximately 10, as the number of times the filter is applied is multiplied by 4 and each application covers more pixels.

Having assessed the performance of the different spherical
shear-based mass aperture implementations in terms of computational time, we now compare their precision.
Assuming that the brute-force version of the galaxy position-based approach is the most precise implementation and cannot be surpassed by any other method, Table~\ref{tab:comp-error-full-cat} compares the three other implementations against this reference implementation in terms of precision. We report both the mean and the standard deviation of the relative difference of the reconstructed E-mode aperture mass maps $M_{\rm ap}^E$ at resolution $\texttt{NSIDE}$ = 2048 and 4096.
This relative difference is defined as follows:

\begin{equation}
\epsilon_{\rm X} =\frac{[M_{\rm ap}^E]^{\rm X} - [M_{\rm ap}^E]^{\rm sphS\_pos}}
  {\sigma_{[M_{\rm ap}^E]^{\rm sphS\_pos}}}
\end{equation}
with $[M_{\rm ap}^E]^{\rm sphS\_pos}$ is the reference E-mode mass aperture map reconstructed using the brute force version of the galaxy position based approach and $[M_{\rm ap}^E]^{\rm X}$ the E-mode mass aperture map reconstructed using another spherical shear-based implementation and $\sigma_{[M_{\rm ap}^E]^{\rm sphS\_pos}}$ is the standard deviation of the reference aperture mass map used to normalise the error, allowing for consistent comparisons across different resolutions.

\begin{table}[t!]
    \centering
    \begin{tabular}{c l S[table-format=1.2e-1] S[table-format=1.2e-1]}
        \toprule
        \textbf{\texttt{NSIDE}} & \textbf{Method} & \textbf{Mean} & \textbf{Standard deviation} \\
        \midrule
        \multirow{3}{*}{2048}
        & $\epsilon_{\rm sphS\_pix}$  & 2.00e-5 & 3.30e-2 \\
        & $\epsilon_{\rm sphS\_pixB}$ & 2.00e-5 & 3.22e-2 \\
        & $\epsilon_{\rm sphS\_posB}$ & 1.54e-6 & 8.02e-3 \\
        \midrule
        \multirow{3}{*}{4096}
        & $\epsilon_{\rm sphS\_pix}$  & 4.29e-6  & 8.58e-3 \\
        & $\epsilon_{\rm sphS\_pixB}$ & 4.36e-6  & 7.51e-3 \\
        & $\epsilon_{\rm sphS\_posB}$ & -3.62e-7 & 3.53e-3 \\
        \bottomrule
    \end{tabular}
    \caption{\footnotesize Comparison of the bias and precision of the relative differences between the spherical shear-based aperture mass implementations presented in Table~\ref{tab:methods} and described in Sect.~\ref{sec:implementation}. The aperture mass map reconstructed using the brute-force galaxy-position-based approach is used as the reference.}
    \label{tab:comp-error-full-cat}
\end{table}

The overall bias and precision are given in Table~\ref{tab:comp-error-full-cat}.
As expected, the test shows that the galaxy-position-based approach with partitioning provides the most precise and accurate reconstruction with respect to the reference implementation. We find that increasing the resolution from \texttt{NSIDE}~=~2048 to 4096 improves the precision of all the methods with a gain by approximately a factor of 2 for the galaxy-position-based approach and a factor of 4 for the pixel-based approaches. The precision of the pixel-based approaches is more strongly dependent on the resolution of the aperture mass maps, which is not surprising. We find that the use of barycentres for the pixel-based approach provides negligible improvement in precision, while remaining a relevant option since it also has a negligible impact on the computational time. In the following sections, we compare these results with those obtained using state-of-the art methods.

\subsection{Cartesian (projected) vs spherical shear-based mass aperture comparison}
\label{sec:cart-vs-sph}
In this subsection, we compare the precision of Cartesian shear-based aperture mass maps, obtained by projecting the galaxy positions and shear onto a Cartesian coordinate system, with that of the spherical shear-based aperture mass implementations developed in this work.

\subsubsection{Cartesian (projected) shear-based implementation}
\label{sec:proj2}
Once the projection is performed properly as explained in Sect.~\ref{sec:proj}, the main challenge in comparing the Cartesian and spherical implementations arises from the different pixel geometries and pixel centres used in each framework. Since the pixel shapes differ, the galaxies assigned to a given pixel are not identical between the two approaches, leading to discrepancies that are independent of the projection method itself. In this work, to isolate the effect of the projection, the comparison has been performed applying the same HEALPix pixelisation scheme to both the Cartesian and spherical analyses. As a result, the shear-based aperture mass maps generated by the two implementations are defined on an identical HEALPix grid, enabling a direct comparison.
Additionally to reduce the errors introduced by the pixelisation or partitioning of the catalogue, the Cartesian (projected) shear-based aperture approach is performed applying the aperture mass
filter directly at the position of the galaxies using a brute force approach.
The algorithm used for the comparison is implemented as follows:
\begin{itemize}
     \item[1.]  The spherical position and shear of galaxies are projected onto a tangent plane centred on a chosen reference point. 
     \item[2.]  The position of the centres of the pixels of the reconstructed HEALPix aperture mass map are also projected onto a tangent plane centred at the same reference point.
     \item[3.]  For each galaxy, the shear is then decomposed into its tangential and cross components, and the filter function $Q$ is evaluated based on the galaxy’s distance from the HEALPix pixel centre. These distances and angular quantities are defined within the Cartesian tangent-plane geometry. 
     \item[4.]  For each HEALPix pixel, the aperture mass and associated noise is then computed by summing the weighted contributions of all galaxies within the aperture, normalising by the mean galaxy density per arcmin$^2$ within the aperture. The latter is obtained by dividing the mean number of galaxies per pixel within the aperture by the pixel area expressed in arcmin$^2$.
\end{itemize}
The proposed algorithm provides the most precise and accurate implementation achievable with respect to the reference implementation based on the Cartesian shear-based approach after tangent projection. 
\subsubsection{Comparison}
\begin{figure}[!h]
   \centering
     \includegraphics[width=\linewidth]{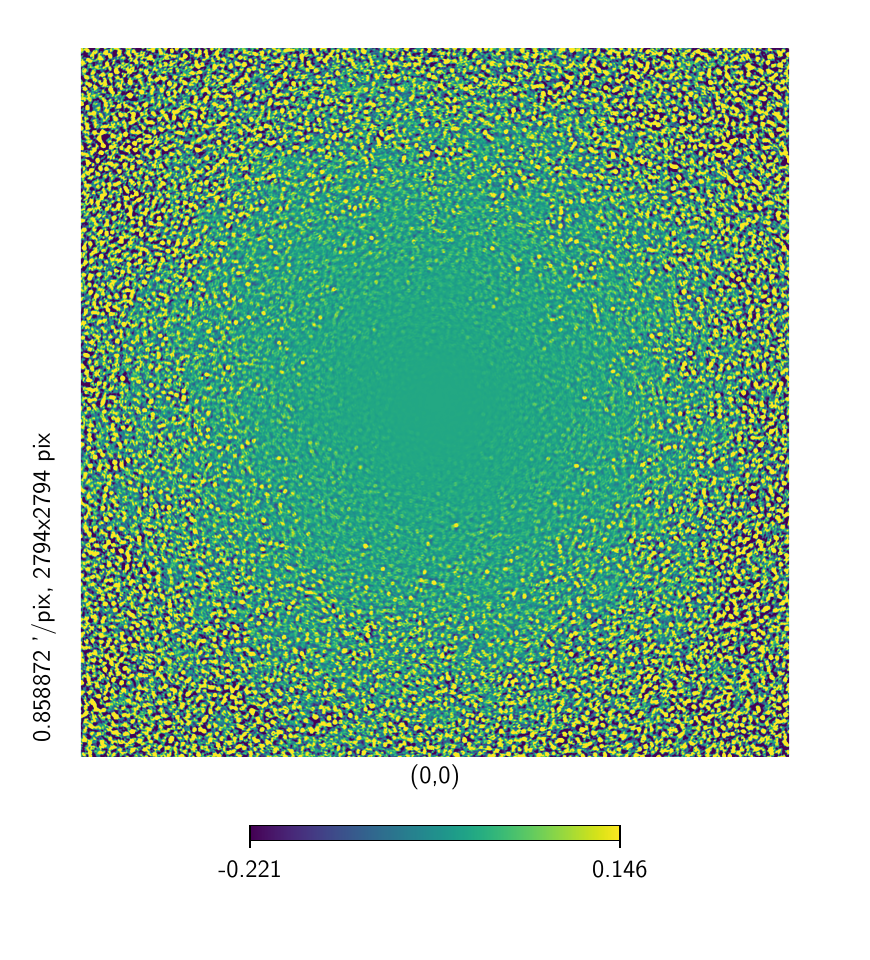}
     \caption{\footnotesize Relative difference map between the Cartesian shear-based aperture mass map, obtained after projecting the galaxy positions and shear onto a Cartesian plane, and the spherical shear-based aperture mass map. Both maps are computed on a HEALPix grid using the brute force galaxy position-based implementation (using the summed approach). The field covers 40$^\circ$~$\times$~40$^\circ$ and is centred on the projection centre. The aperture mass maps are computed using a resolution of \texttt{NSIDE}~=~4096 and following the summed approach.}
     \label{fig:sauron}
\end{figure}

Figure~\ref{fig:sauron} shows the relative difference between the E-mode aperture mass maps reconstructed using the Cartesian formalism after projecting the galaxy positions and shear onto a Cartesian plane described in Sect.~\ref{sec:proj2} and the spherical formalism using the reference implementation described in Sect.~\ref{sec:full-cat}. Both maps are computed on a HEALPix grid for a resolution of \texttt{NSIDE}~=~4096 using the brute force galaxy position-based implementation. 
The relative difference map shown in Fig.~\ref{fig:sauron} covers a field of view of 40$^\circ$~$\times$~40$^\circ$. While the flat-sky approximation remains valid for small sky patches, the small-angle approximation becomes increasingly inaccurate for fields larger than 10$^\circ$, leading to significant distortions at increasing distances from the projection centre.

\begin{figure}[!h]
   \centering
     \includegraphics[width=\linewidth]{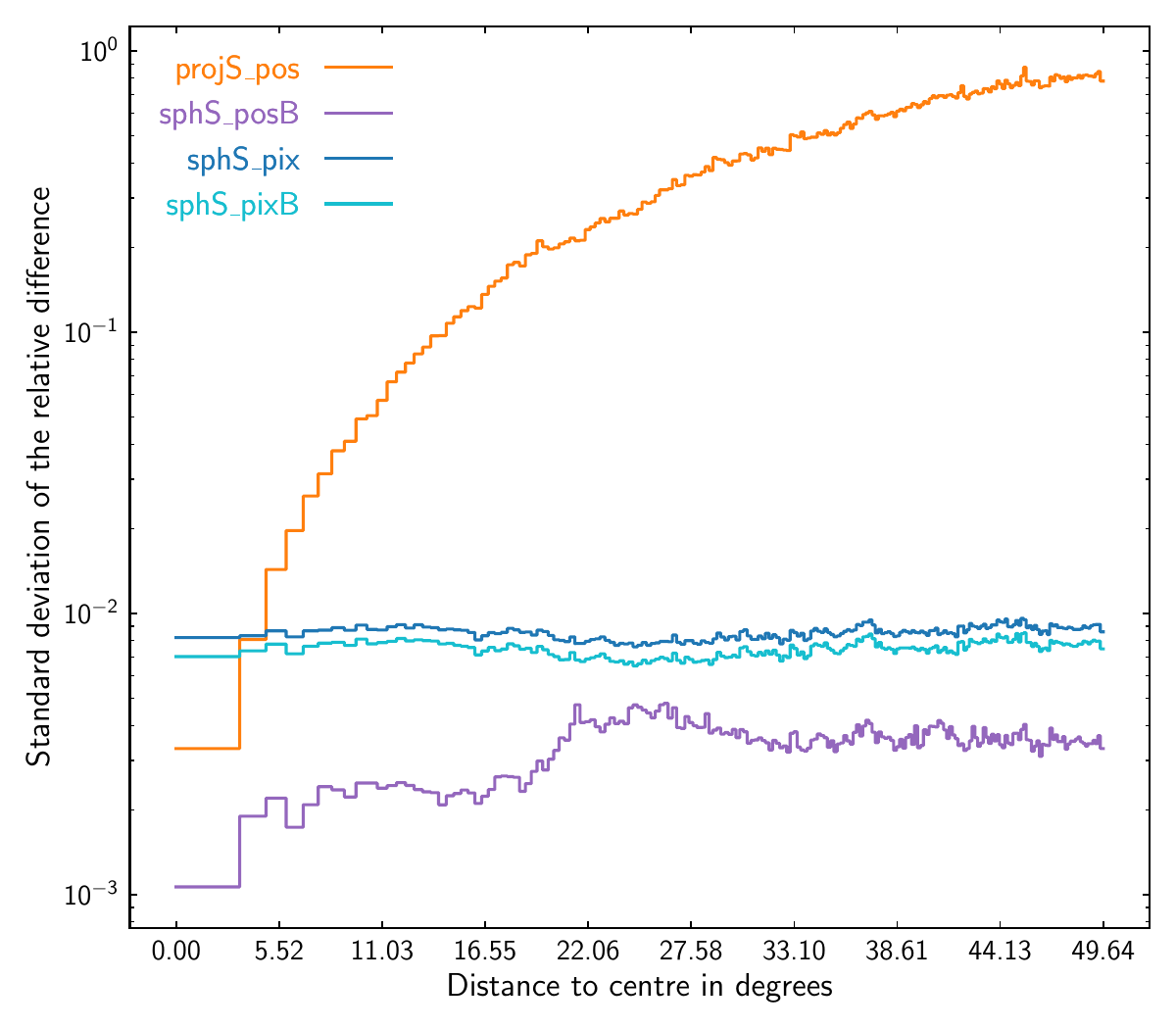}
     \caption{\footnotesize Standard deviation of the relative difference as a function of the distance to the centre of the Cartesian projection, comparing the optimised spherical shear-based : pixel-based (blue), pixel-based using barycentre (cyan) and galaxy position-based using HEALPix partitioning (purple) and the Cartesian (projected) shear-based (orange) mass aperture implementations to the spherical brute force implementation. The standard deviation is computed within annular bins of equal area. The aperture mass maps are computed using a resolution of \texttt{NSIDE}~=~4096 and following the summed approach.}
     \label{fig:cart_vs_sph}
\end{figure}

To provide a more quantitative assessment of the errors introduced by the flat-sky approximation and to compare them with those arising from the optimised spherical shear-based implementations presented in this work, Fig.~\ref{fig:cart_vs_sph} presents a comparison of the errors introduced by these different implementations. The three optimised spherical shear-based implementations described in Sect.~\ref{sec:implementation}, as well as the Cartesian (projected) shear-based implementation described in Sect.~\ref{sec:proj2}, are evaluated against the reference shear-based implementation. The figure shows the evolution of the standard deviation of the relative difference as a function of the distance from the centre of the Cartesian projection at a resolution of \texttt{NSIDE}~=~4096. The standard deviation is computed within annular bins of equal area.

As expected, and in agreement with the results presented in Table~\ref{tab:comp-error-full-cat}, the most precise implementation with respect to the reference is the galaxy-position-based approach with HEALPix catalogue partitioning. The observed variation of the error with the distance from the projection centre is due to the fact that the HEALPix pixels used for the partitioning are not uniformly distributed over the sphere. Since the projection centre is located on the equator, integrating over concentric annuli around this point leads to an increase in the error towards the poles. As expected for the Cartesian (projected) shear-based approach, the relative difference increases with distance from the projection centre. This evolution is due to the limitations of the flat-sky approximation. At a resolution of \texttt{NSIDE}~=~4096 and at a distance of 5$^\circ$ from the centre of the projection, all the spherical shear-based implementations become more precise than the Cartesian approach after projection. 

These results show that spherical shear-based aperture mass implementations should be preferred in terms of precision. Moreover, in terms of computational time, the Cartesian (projected) shear-based approach has a runtime comparable to that of the reference spherical shear-based brute-force implementation, while achieving lower precision. Even the spherical shear-based approach using pixelisation provides better performance both in terms of precision and computational time.

\subsection{Shear vs convergence-based mass aperture comparison}
\label{sec:comp_conv}


In this section, we aim to compare the precision between the spherical convergence-based aperture mass approach and the different shear-based implementations presented in Table~\ref{tab:methods}. As explained in Sect.~\ref{sec:SUMvsAVG}, the spherical convergence-based aperture mass maps are inherently computed using the averaged approach, meaning that they are normalised by the effective galaxy density per arcmin$^2$ within the pixel.
In this section only, to enable a consistent comparison, we also adopt the averaged approach when computing both the spherical shear-based aperture mass maps and the Cartesian (projected) shear-based aperture mass maps.


\subsubsection{Convergence-based implementation}
\label{sec:conv2}
The algorithm used to produce the spherical convergence-based aperture mass maps is implemented as follows:

\begin{itemize}
     \item[1.]  The shear maps are constructed by discretizing the full sphere into a regular HEALPix grid. The spherical position and shear of galaxies are then projected onto this grid.
     \item[2.]  For each grid pixel, the shear component $\gamma_1$ and $\gamma_1$ are computed by summing the corresponding galaxy shear components $\gamma_1$ and $\gamma_2$ over all the galaxies contained within the pixel. The resulting summed shear components are then normalised by the number of galaxies in the pixel and multiplied by the pixel area expressed in arcmin$^2$.
     \item[3.]  The spherical convergence maps are constructed from these shear maps using the spherical extension of the standard mass inversion formalism introduced by \cite{Kaiser_1993} \cite[e.g.][]{Chang_2018}. 
     \item[4.]  The value of the filter $U$ at the position of the pixels within the aperture is calculated from its distance to the pixel centre.
     \item[5.]  The aperture mass associated with a given HEALPix pixel is then computed by summing the weighted contributions of all galaxies within the aperture.
\end{itemize}

As noted earlier, a limitation of the spherical convergence-based implementation is that, unlike the shear-based implementations, it does not provide a direct estimate of the associated noise.


\subsubsection{Comparison}

\begin{figure}[!t]
   \centering
     \includegraphics[width=\linewidth]{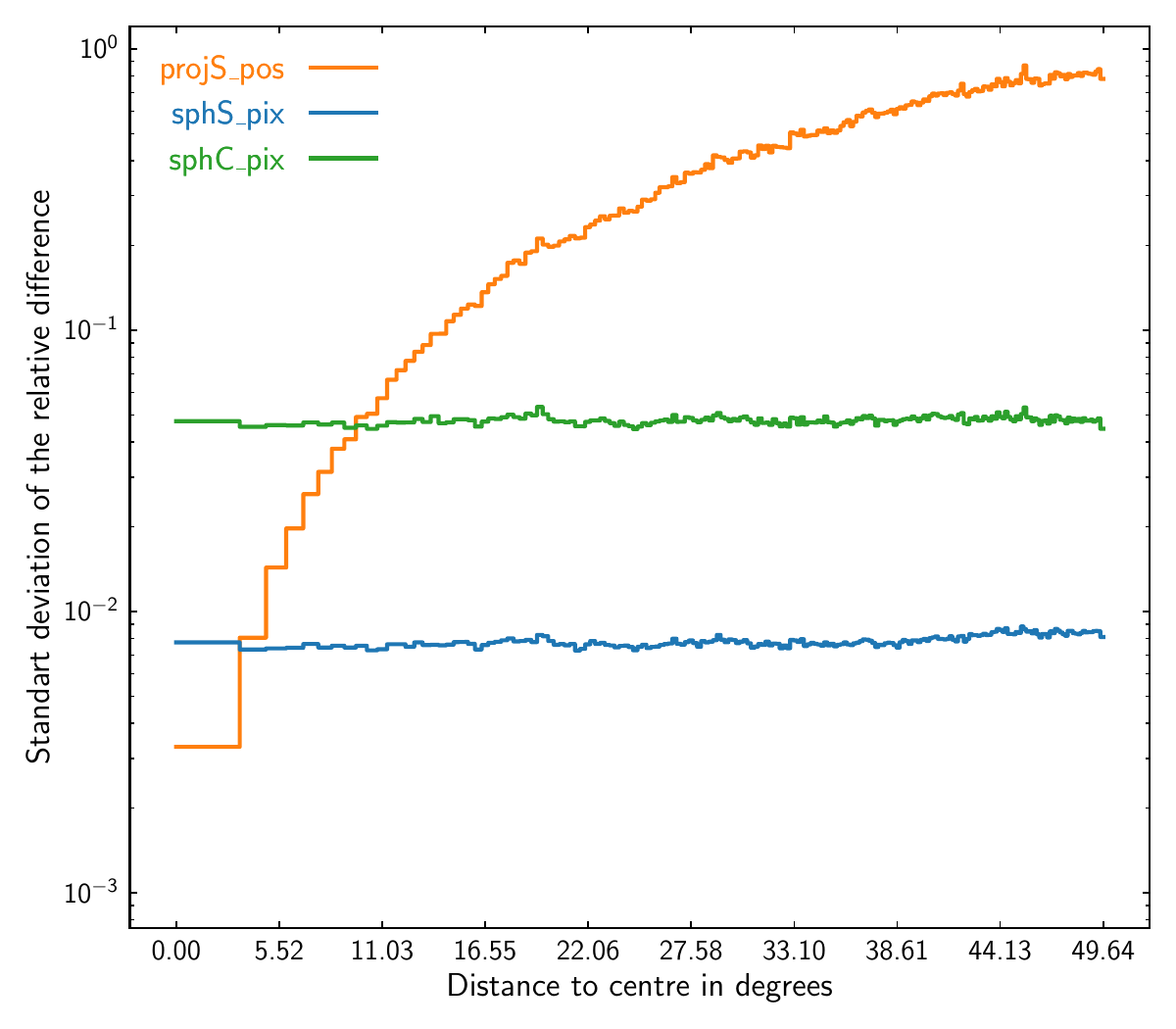}
     \caption{\footnotesize Standard deviation of the relative difference as a function of the distance to the centre of the Cartesian projection, comparing the spherical convergence-based (green), the spherical shear-based (blue) and the Cartesian (projected) shear-based (orange) mass aperture implementations to the spherical brute force implementation. The standard deviation is computed within annular bins of equal area.  The aperture mass maps are computed using a resolution of \texttt{NSIDE}~=~4096 and following the averaged approach.}
     \label{fig:conv_vs_sph}
\end{figure}

Figure~\ref{fig:conv_vs_sph} compares the spherical convergence-based implementation, the spherical shear-based implementation using the pixel-based approach described in Sect.~\ref{sec:pix-map} (the least precise of the spherical shear-based approaches), and the Cartesian (projected) shear-based implementation described in Sect.~\ref{sec:proj2} against the reference shear-based implementation. The figure shows the evolution of the standard deviation of the relative difference as a function of the distance from the centre of the Cartesian projection at a resolution of \texttt{NSIDE}~=~4096. The standard deviation is computed within annular bins of equal area. In Fig.~\ref{fig:conv_vs_sph}, we find that the precision of the spherical convergence-based implementation is approximately one order of magnitude lower than that of the least precise spherical shear-based implementation. Furthermore, this comparison is carried out under ideal conditions, without any survey mask. In realistic survey conditions, the presence of a mask would introduce additional systematic effects in the convergence-based approach unless they were explicitly corrected.

\label{sec:comp2}
In terms of computational time, the spherical convergence-based approach has a runtime two times faster than the spherical shear-based implementation using pixelisation. However, the time to compute the spherical convergence-based aperture mass map is multiplied by at least a factor of 100, if we use the mass inversion method proposed in \cite{Pires_2020} that includes corrections for systematic effects. 
Additionally, as discussed in Sect.~\ref{sec:conv}, applying the aperture mass filter to convergence maps does not provide a straightforward way to estimate the associated noise. Instead, the noise in the aperture mass maps must be estimated by applying the aperture mass filter to $N$ convergence maps reconstructed from catalogues in which the source galaxy orientations have been randomly rotated. This is computationally costly and becomes particularly prohibitive when using more sophisticated mass inversion methods.
These results demonstrate that spherical shear-based aperture mass implementations should be preferred to spherical convergence-based aperture mass implementation. In particular, even the pixel-based spherical shear-based implementation outperforms the spherical convergence-based approach in terms of both precision and computational time.



\label{sec:shear-vs-cov}
\section{Application to Weak Lensing cluster detection}
\label{sec:application}
In this section, we assess the performance of the spherical shear-based implementations for weak lensing cluster detection. 
The weak lensing cluster detection is performed using the same galaxy catalogue as in the previous section with a non-uniform galaxy density of
27 gal/arcmin$^2$ and covering an area of 7372.47 deg$^2$. Shape noise, arising from the intrinsic ellipticity distribution of galaxies and measurement errors, was modelled as Gaussian noise with zero mean and dispersion $\sigma_{\epsilon} = 0.26$ per shear component \cite[e.g.][]{Leauthaud_2007}.

\subsection{Detection procedure}
The detection procedure starts by computing the spherical shear-based aperture mass maps. As in Sect.\ref{sec:evaluation} we adopt the aperture mass filter function proposed in \cite{Jarvis_2004}, which targets objects at a scale of 4'. We test both the least precise spherical shear-based implementation and the Cartesian (projected) shear-based implementation. In line with the results of \citep{Leroy_2023}, we adopt a summed approach that is better suited for weak lensing cluster detection. As explained in the previous section, the spherical convergence-based implementation is not considered in this application because the estimation of the noise is prohibitive.

Then, pixels in the aperture mass maps with values higher than their eight direct neighbours were identified as detections. 
In order to assess the purity and completeness, the detections are matched with the halos in the catalogue depending on the angular distance between each detection and the closest catalogue halo, following the matching procedure described in \cite{Chappuis_2026}.



\subsection{Results}
\begin{figure}[!h]
   \centering
     \includegraphics[width=\linewidth]{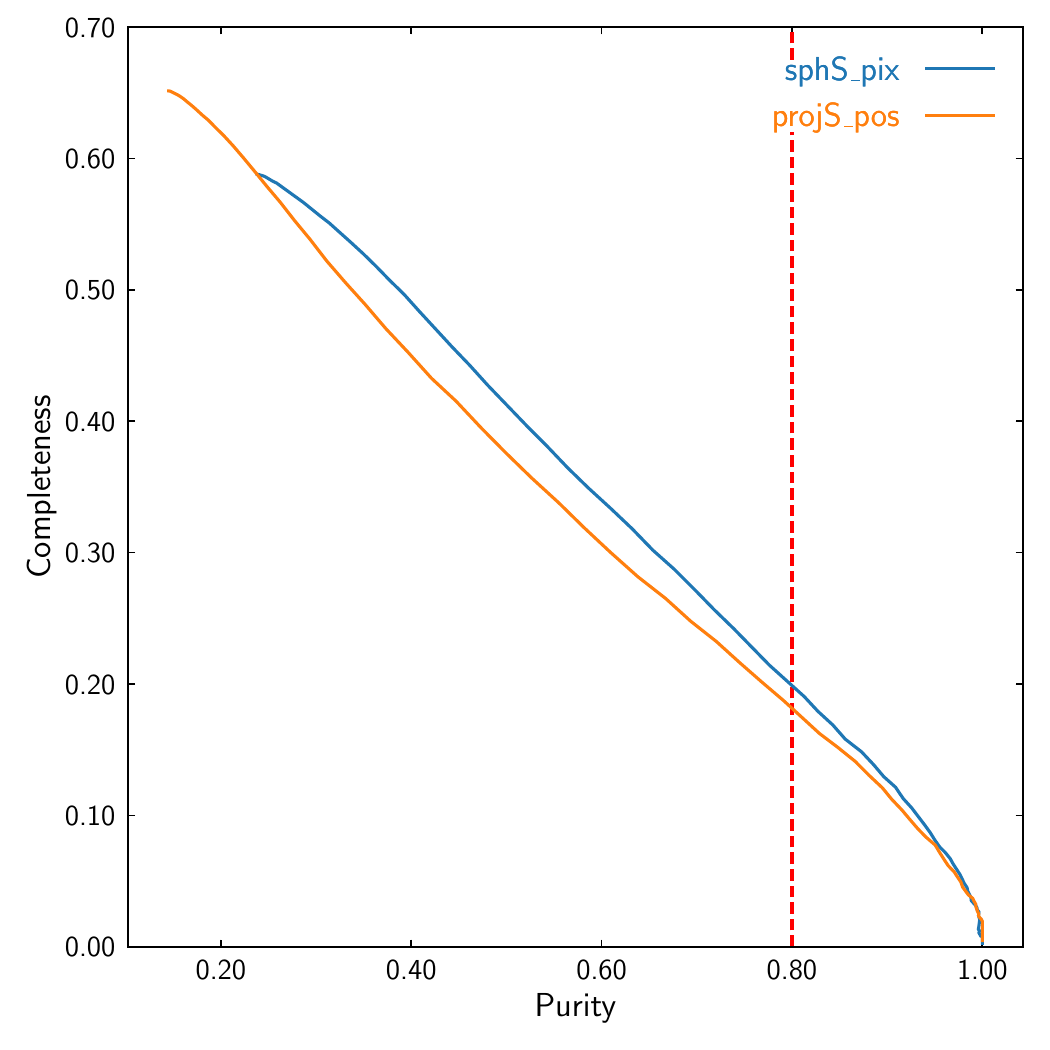}
    \caption{\footnotesize Completeness as a function of purity for the detection method  based on the spherical shear-based aperture mass implementation (blue) and the Cartesian (projected) shear-based aperture mass implementation (orange).}
     \label{fig:purityvscomp}
\end{figure}

Fig.\ref{fig:purityvscomp} shows the evolution of the completeness as a function of the purity when varying the threshold in the detection procedure. We see that the spherical shear-based consistently outperforms the Cartesian (projected) shear-based aperture mass implementation achieving higher completeness across the full range of purity values. At a purity level of 80\%, the spherical implementation provides an increase in completeness of approximately 10\%.

\begin{figure}[!h]
   \centering 
    \includegraphics[width=\linewidth]{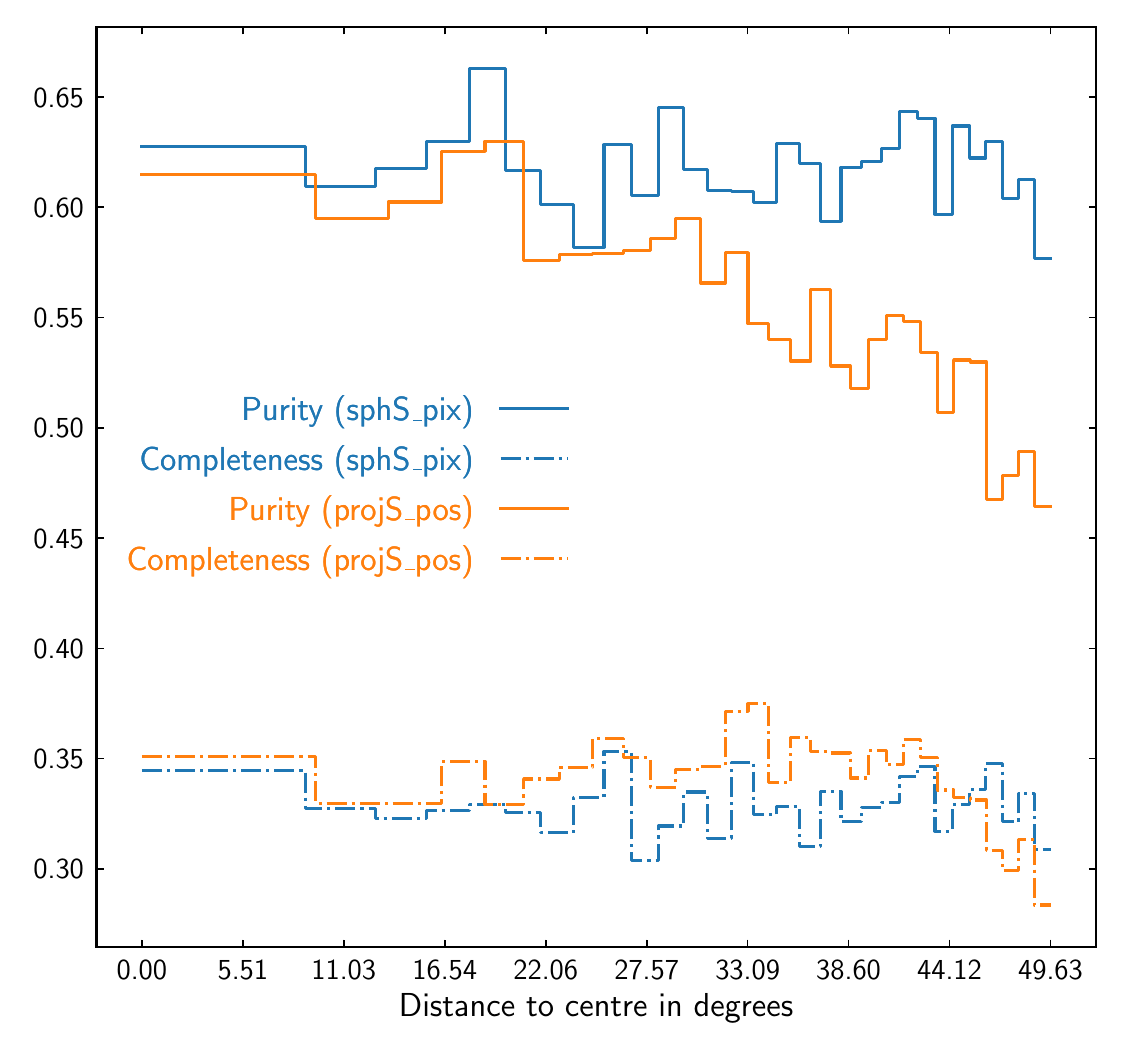}
    \caption{\footnotesize Evolution of the purity (solid line) and completeness (dashdotted line) as a function of the distance to the centre of the projection for the detection method based on the spherical shear-based aperture mass implementation (blue) and the Cartesian (projected) shear-based aperture mass implementation (orange).}
    \label{fig:purityandcomp}
    \end{figure}

To better quantify the impact of the errors introduced by the flat-sky approximation, we fix a common detection threshold for both detection methods and evaluate the resulting purity and completeness as functions of the distance from the projection centre. Fig.\ref{fig:purityandcomp} shows the evolution for the detection method based on the spherical shear-based aperture mass implementation (blue) and the Cartesian (projected) shear-based aperture mass implementation (orange).
We find that the completeness of the two detection methods is quite similar, although the Cartesian shear-based implementation shows a small decrease at large distances from the projection centre. As expected, the purity remains constant for the spherical shear-based implementation. Conversely, the purity of the Cartesian (projected) shear-based implementation decreases with increasing distance from the projection centre, indicating that the flat-sky approximation becomes progressively inaccurate away from the projection centre.

These results confirm our previous findings that the spherical shear-based implementation improves the precision of the mass aperture reconstruction compared to state-of-the-art method, in particular by increasing the number of confirmed galaxy cluster detections. We show that even the least precise spherical shear-based implementation proposed in this work (using pixel-based approach) increases the completeness by a factor of 10\% compared to the Cartesian (projected) shear-based implementation.

Another issue of the Cartesian (projected) shear-based approach that is particularly problematic for cluster detection arises from projection effects. Projecting the spherical shear field onto a tangent plane distorts the effective distances between points away from the projection centre, which results in an artificial increase in angular distances with increasing separation from the projection centre. As a consequence, for a fixed filter size defined in the projected plane, the effective angular scale of the filter decreases with increasing distance from the projection centre. Therefore, the filter does not probe the same angular cluster scales across the field, resulting in a position-dependent selection of clusters.

\section{Conclusions}
\label{sec:conclusion}
The increasing size of weak lensing surveys requires a correct treatment of weak lensing measurements directly on the celestial sphere. While the spherical formalism has been derived to reconstruct the convergence maps from the shear field, the derivation of aperture mass maps from the shear field has been lacking until now. These spherical shear-based mass aperture methods are particularly suited for cluster studies, as they enable the detection of clusters at specific scales. They are also valuable for cosmological analyses based on higher-order statistics such as peak counts, moments, and probability density functions estimated at different scales.

In this paper, we have introduced the spherical formalism for the computation of mass aperture maps directly from the shear field defined on the celestial sphere. To this end, we introduce and compare four different implementations, ranging in precision and computation time. The codes were validated using a simulated galaxy catalogue with a realistic spatial galaxy distribution. This catalogue was specifically developed in this work to provide a realistic framework to evaluate spherical mass aperture in the context of weak-lensing cluster detection. Unlike existing approaches, our method does not rely on planar projections and does not require the intermediate reconstruction of convergence maps.

All the proposed implementations achieve better precision than existing approaches. In particular, we find that, at a resolution of $N_{\rm SIDE}=4096$ and at $5^\circ$ from the centre of the projection, all of the proposed spherical shear-based implementations achieve higher precision than the standard approach based on planar projection of the shear. The gain in precision increases further with distance from the projection centre. At 10°, we observe an improvement of approximately one order of magnitude. This is because the error associated with the planar projection increases with increasing distance from the projection centre, as a consequence of the limitations of the flat-sky approximation.
 We also find that, under ideal conditions without masks, the spherical convergence-based implementation is approximately one order of magnitude less precise than the least precise spherical shear-based implementation, in addition to not allowing for direct estimation of the corresponding noise maps.
 The comparison further demonstrates that the two proposed spherical shear-based implementations using a pixel-based approach outperform the existing methods in terms of both precision and computational time.
 Depending on the application and the number of times the code must be executed, one of the proposed implementations may be preferred over the other as a trade-off between computational efficiency and precision.

In this paper, we also implemented two different pixel normalisation schemes -- summed and averaged -- showing that they achieve similar execution time and precision. Therefore, the choice between the two should only be guided by the specific application.

Finally, we showed that the spherical shear-based aperture mass implementations improve the quality of cluster detection through the weak lensing effect. In particular, we showed that the cluster sample produced using the spherical shear-based implementation consistently outperforms that  obtained using projected shear-based implementation in the purity–completeness plane.
In particular, for a given detection threshold, we find that the purity obtained with the projected method decreases with increasing distance from the projection centre. This result is expected, as the flat-sky approximation becomes progressively less accurate as the distance from the projection centre increases. Additionally, projecting the spherical shear field onto a tangent plane distorts the effective distances between points particularly as their separation from the projection centre increases. For weak-lensing cluster detection, this projection effect modifies the effective size of the aperture mass filter and introduces a position dependence in the cluster selection, which is particularly problematic for modelling the selection function. These results demonstrate the importance of accounting for the spherical geometry in weak-lensing cluster detection, both to ensure a uniform selection across the sky and to enable a consistent modelling of the resulting cluster sample.
The spherical mass aperture software presented in this paper, the Spherical Mass Aperture Toolkit (SMART) is publicly available on \href{https://github.com/cea-lilas/SMART}{GitHub}.

\begin{acknowledgements}
The authors used ChatGPT (OpenAI) to improve the language and readability of parts of the manuscript. The authors reviewed and edited the generated suggestions and take full responsibility for the content of the manuscript.
\end{acknowledgements}

\bibliographystyle{aa} 
\bibliography{biblio} 
\appendix

\end{document}